# Study of the Magnetic Properties of a Lieb Core-Shell Nano-Structure: Monte Carlo Simulations

S. Aouini [1], S. Ziti [2], H. Labrim [3,*] and L. Bahmad [1,*]

1 Laboratoire de la Matière Condensée et Sciences Interdisciplinaires LaMCScI, Mohammed V, University of Rabat, Faculty of Sciences, B.P. 1014, Rabat, Morocco

[2] Laboratoire de Recherche en Informatique, Mohammed V University in Rabat, Faculty of Sciences, B.P. 1014, Rabat, Morocco

[3] Centre National de l'Energie, des Sciences et des Techniques Nucléaires (CNESTEN), Rabat, Morocco

**Abstract:**

The object of this work is to study the magnetic properties of the Lieb structure based on magnetic particles with «core-shell» nano-structure. This system is consisting of a ferromagnetic 2D Ising model formed with spins: $\sigma = 1/2$, $S = 1$ and $q = 3/2$. Firstly, we examine the magnetic properties of this nanostructure. On the other hand, we study the effect of the coupling exchange interactions and the crystal field. We use the Monte Carlo simulations to investigate also the behavior of the Lieb nano-structure. Finally, we study the influence of the temperature on the critical and compensation behaviors of the studied system. The obtained results are summarized in the phase diagrams.

[*] Corresponding authors : bahmad@fsr.ac.ma; lahou2002@gmail.com

## 1- Introduction :

Nowadays, the various nano-structures have become important elements of the science and industrial research. Geometry, whether on the atomic or nano-scale, is a key factor for the electronic band structure of materials. Some specific geometry gives rise to novel and potentially useful electronic bands. For example, a honeycomb lattice leads to Dirac type bands where the charge carriers behave as mass-less particles [1]. A significant advantage of the scientific research is the full control of a wide range of system parameters as, for example, lattice geometry, interaction, and disorder. The design of the molecular Lieb lattice is not trivial for several reasons. When, the Lieb lattice has four-fold rotational symmetry.

Theoretical predictions are triggering the exploration of novel two-dimensional (2D) geometries [2-10], such as graphynes and the kagomé and Lieb lattices. The Lieb lattice is the 2D analogue of the 3D lattice exhibited by perovskites [2]; it is a square-depleted lattice, which is characterized by a band structure featuring Dirac cones intersected by a flat band. Whereas photonic and cold-atom Lieb lattices have been demonstrated [11-17], an electronic equivalent in 2D is difficult to realize in an existing material. On the other hand, the heat capacity of the magnetic susceptibility reduces with increasing the magnetic field strength in the high temperature region of the spin Lieb nano-lattice [18].

Since a 2D electronic Lieb lattice has not been realized, lithography could be used to impose a Lieb pattern on a 2D electron gas [7, 19]. Alternatively, a similar strategy is envisaged to the one employed for generating artificial grapheme and could be used for the assembling a molecular lattice on a substrate featuring a surface state to force the electrons into the desired geometry [20].

The aim of this paper is to examine the magnetic properties of the Lieb magnetic composites based on magnetic particles with «core-shell» nano-structure consisting of a ferromagnetic in 2D Ising model. We study the effect of the coupling exchange interactions and the crystal field. Indeed, we analyze the influence of the temperature on the critical and compensation behaviors of the studied system by using the Monte Carlo method [21-24].

The outline of this paper is as follows: We describe the model and the formulations used and method in section 2. Section 3 is dedicated to results and discussions. A conclusion is in given in section 4.

## 2- Model and method:

In this section, we study the magnetic properties of a Lieb nano-structure compound based on magnetic spins: $\sigma = 1/2$, $S = 1$ and $q = 3/2$. Such structure consists of a ferromagnetic 2D Ising model. We examine the influence of the critical and compensation temperature behaviors under a uniform magnetic field. We examine also the effect of the crystal field and the coupling exchange interactions. The corresponding geometry of the studied system is given in Fig. 1.

The Hamiltonian controlling the studied Lieb nano-structure is given by:

$$\mathcal{H} = -J_{\sigma\sigma} \sum_{i,j} \sigma_i \sigma_j - J_{ss} \sum_{k,l} S_k S_l - J_{qq} \sum_{m,n} q_m q_n - J_{\sigma s} \sum_{i,k} \sigma_i S_k - J_{\sigma q} \sum_{i,m} \sigma_i q_m - H \sum_i (S_i + \sigma_i + q_i) - D_S \sum_i S_i^2 - D_q \sum_i q_i^2 \qquad (1)$$

For simplicity, $D_S$ and $D_q$ are crystal fields corresponding to the magnetic atoms S and q. We will limit our study to the special case: $D = D_S = D_q$.

- $J_{\sigma\sigma}$, $J_{ss}$ and $J_{qq}$ are the coupling exchange interactions constant between the two first nearest neighbor atoms with spins $\sigma - \sigma$, $S - S$ and $q - q$, respectively.
- $J_{\sigma s}$ is the coupling exchange interaction constant between two nearest neighbor magnetic atoms with spin $\sigma - S$.
- $J_{\sigma q}$ stands for the coupling exchange interaction constant between two nearest neighbor magnetic atoms with spin $\sigma - q$.
- The crystal field D is applied on the S-spins and q-spins. Where: $D = D_s = D_q$.
- H is the external longitudinal magnetic field.

The number of spins of the studied system is: $N_\sigma = 5$, $N_S = 4$ and $N_q = 12$. The total number spins is $N = N_\sigma + N_S + N_q = 21$, see Fig. 1.

We apply the Monte Carlo under the Metropolis algorithm [25-29] to inspect the magnetic behavior of the studied Lieb magnetic «core-shell» nano-structure. The system is formed with the spins (1/2, 1, 3/2). During the calculations we generate a random initial configuration. This method has been performed to investigate the equilibrium nanosystem [30]. We perform $10^6$ MC steps per spin by visiting the sites of the system.

The steps per spin MCS reaches the equilibrium system for $10^5$ Monte Carlo steps, for a fixed temperature. We eliminate the first $10^4$ configurations for each initial condition. The error bars were calculated with a Jackknife method [31].

The internal energy per site is:

$$E = \frac{1}{N}\langle \mathcal{H} \rangle \qquad (2)$$

The magnetizations per are defined by:

$$m_\sigma = \frac{\sum_{n=1}^{N_S} \sigma_n}{N_S} \qquad (3)$$

$$m_S = \frac{\sum_{n=1}^{N_S} S_n}{N_S} \qquad (4)$$

$$m_q = \frac{\sum_{n=1}^{N_q} q_n}{N_q} \qquad (5)$$

With: $N_\sigma = 5$, $N_S = 4$ and $N_q = 12$.

At the compensation point, the total magnetization undergoes it null value, when the condition $|M_{shell}(T_{comp})| = |M_{core}(T_{comp})|$ is verified.

The total magnetic susceptibility $\chi$ is given by:

$$\chi = \beta N(<M^2> - <M>^2) \qquad (6)$$

Where, N is the total number of spins:

$$N = N_\sigma + N_S + N_q \qquad (7)$$

The total magnetization is expressed as:

$$M = \frac{\sum_1^N N_\sigma M_\sigma + N_S M_S + N_q M_q}{N} \qquad (8)$$

Where: $\beta = 1/k_B T$, where T is the absolute temperature and ($k_B = 1$) is Boltzmann's constant.

3- **Results and discussion:**

   3-1- **Ground state phase diagrams:**

Starting from the Hamiltonian (1) describing the studied system, we inspect the possible stable configurations in different phase planes of the physical system. By computing and comparing the all possible configurations $2 \times 3 \times 5 = 30$, we provide the corresponding phase diagrams in different planes. Indeed, we combine the different configurations of the spins $\sigma = 1/2$, $S = 1$ and $q = 3/2$. The corresponding ground state phase diagrams are presented in Figs. 2(a)-(k).

In the plane (H, D), we illustrate in Fig. 2a, the stable phases for the exchange coupling interactions values: $J_{\sigma\sigma} = J_{ss} = J_{qq} = J_{\sigma s} = J_{\sigma q} = 1$. We found that six stable configurations are present in this figure, for the negative values of the crystal field D. While, only two configurations are found to be stable for the positive values of this parameter, see the Fig. 2a. To inspect the stable configurations, we illustrate in Fig. 2b, corresponding to the plane (H, $J_{ss}$), the stable phases, for the exchange coupling interactions values: $J_{\sigma\sigma} = J_{qq} = J_{\sigma s} = J_{\sigma q} = 1$. From this figure, a perfect symmetry is found, regarding the external magnetic field H for (D = 0), see the Fig. 2b. We found that four stable configurations are present, for the negative values of the coupling exchange interaction $J_{ss}$. Only two configurations are found to be stable for the positive values of this parameter.

A perfect symmetry is found, in Fig. 2c, regarding the external magnetic field H in the plane (H, $J_{\sigma s}$) for the exchange coupling interactions fixed to the values: $J_{\sigma\sigma} = J_{ss} = J_{qq} = J_{\sigma q} = 1$. In the absence of the crystal field (D = 0), we found that four stable configurations, corresponding to negative values of the coupling exchange interaction $J_{\sigma s}$. In fact, only two configurations are found to be stable for the positive values of this parameter.

In the plane (D, $J_{ss}$), corresponding to Fig. 2d, we provide the stable configurations for fixed values of the exchange coupling interactions constant : $J_{\sigma\sigma} = J_{qq} = J_{\sigma s} = J_{\sigma q} = 1$. From this figure, we found that four configurations are stable for negative values of both the crystal field D and the coupling exchange interaction $J_{ss}$. Nevertheless, only two configurations are stable for the positive values of the crystal field D. In addition, three configurations are found to be stable for the positive values of the coupling exchange interaction $J_{ss}$ in the absence of the external magnetic field (H = 0), see the Fig. 2d.

In the absence of the external magnetic field (H = 0) in the plane (D, $J_{\sigma s}$), we found Fig. 2e that five configurations stable for negative values of both the crystal field D and the coupling exchange interaction $J_{\sigma s}$. The other exchange coupling interactions are kept constant, in this figure: $J_{\sigma\sigma} = J_{ss} = J_{qq} = J_{\sigma q} = 1$. Indeed, only two configurations are found to be stable for the positive values of the crystal field D and three configurations are found to be stable for the positive values of the coupling exchange interaction $J_{\sigma s}$. This is due to the competition between the crystal field and the exchange coupling interactions effect.

To inspect the stable configurations, in the plane (D, $J_{\sigma q}$) we provide the obtained results in the Fig. 2f, for fixed values of the coupling exchange interactions constant: $J_{\sigma\sigma} = J_{ss} = J_{qq} = J_{\sigma s} = 1$. We found that five configurations are stable, for the negative values of both the crystal field D and the coupling exchange interaction $J_{\sigma q}$. From this figure, two configurations are found to be stable for the positive values of the crystal field D and three configurations are found to be stable for the positive values of the coupling exchange interaction $J_{\sigma q}$ in the absence of the external magnetic field (H = 0).

In Fig. 2g, we present the five stable configurations in the plane ($J_{\sigma\sigma}$, $J_{ss}$), in the absence of the external magnetic and the crystal fields (H = D = 0), for the negative values of the coupling exchange interactions $J_{\sigma\sigma}$ and $J_{ss}$ and fixed values of the exchange coupling interactions: $J_{qq} = J_{\sigma s} = J_{\sigma q} = 1$. From this figure, only two configurations are found to be stable for the positive values of the coupling exchange interactions $J_{\sigma\sigma}$ and $J_{ss}$, see Fig. 2g.

We found that three configurations are stable for the negative values of the coupling exchange interactions $J_{\sigma\sigma}$ and $J_{\sigma s}$, in the plane ($J_{\sigma\sigma}$, $J_{\sigma s}$) for fixed values of the coupling exchange interactions: $J_{ss} = J_{qq} = J_{\sigma q} = 1$, in the absence of both the external magnetic and the crystal fields (H = D = 0). Two configurations are found to be stable for the positive values of the coupling exchange interactions $J_{\sigma\sigma}$ and $J_{\sigma s}$, see the Fig. 2h.

In the absence of the external magnetic and the crystal fields (H = D = 0), we found that four stable configurations are present in the plane ($J_{\sigma s}$, $J_{ss}$), for the negative values of the coupling exchange interactions $J_{\sigma s}$ and $J_{ss}$ for the fixed values of the coupling exchange interactions: $J_{\sigma\sigma} = J_{qq} = J_{\sigma q} = 1$. Moreover, only two configurations are found to be stable for the positive values of the coupling exchange interactions $J_{\sigma s}$ and $J_{ss}$, see the Fig. 2i.

In the plane ($J_{\sigma q}$, $J_{ss}$) we found that four stable configurations are present for the negative values of the coupling exchange interactions $J_{\sigma q}$ and $J_{ss}$ in the absence of both the external magnetic and the crystal fields (H = D = 0). While, only two configurations are found to be stable for the positive values of the coupling exchange interactions $J_{\sigma q}$ and $J_{ss}$, see the Fig. 2j.

In the absence of both the external magnetic and the crystal fields (H = D = 0) in the plane ($J_{\sigma q}, J_{\sigma s}$), we found that only four configurations are stable for the negative values of the coupling exchange interactions $J_{\sigma q}$ and $J_{\sigma s}$ when fixing the values of the coupling exchange interactions: $J_{\sigma\sigma} = J_{ss} = J_{qq} = 1$. On the other hand, only two configurations are found to be stable for the positive values of the coupling exchange interactions $J_{\sigma q}$ and $J_{\sigma s}$, see the Fig. 2k.

### 3-2- Monte Carlo study:

In this part, we use the Monte Carlo simulations to study the magnetizations and the susceptibilities as a function of both of the external magnetic and the crystal fields. We will also examine the effect of the coupling exchange interactions for fixed values of the temperature. The corresponding phase diagrams are plotted in Figs. 3(a)-(d) for a fixed temperature value: T = 0.5 K.

The Fig. 3a presents the behavior of the magnetizations as a function of the external magnetic field in the absence of the crystal field (D = 0) and fixed values of the coupling exchange interactions: $J_{\sigma\sigma} = J_{ss} = J_{qq} = 1$ and $J_{\sigma s} = J_{\sigma q} = -0.1$. The magnetization $m_\sigma$ is of second order transition type; while the magnetizations $m_s, m_q$ and $m_T$ are of first order transition type. Indeed, in the corresponding hysteresis cycles we found a double loop for the total magnetization $m_T$. This behavior can be explained by the anti-ferromagnetism introduced by the negative values of the coupling exchange interactions ($J_{\sigma s} = J_{\sigma q} = -0.1$).

In order to investigate the behavior of the susceptibilities as a function of the external magnetic field in the absence of the crystal field (D = 0) we plot in Fig. 3b the obtained results. This figure is plotted for fixed values of the coupling exchange interactions: $J_{\sigma\sigma} = J_{ss} = J_{qq} = 1$ and $J_{\sigma s} = J_{\sigma q} = -0.1$. In fact, the susceptibilities $\chi_s, \chi_q$ and $\chi_T$ present a peak surrounding the critical temperature value.

Following the same motivation, we illustrate in Fig. 3c the behavior of the magnetizations as a function of the crystal field for (H = 0) and fixed values of the coupling exchange interactions: $J_{\sigma\sigma} = J_{ss} = J_{qq} = 1$ and $J_{\sigma s} = J_{\sigma q} = -0.1$. The magnetization $m_\sigma$ is found to be of second order transition type. While the magnetizations $m_s, m_q$ and $m_T$ are of first order transition type. These behaviors are started from the negative values to the positive ones. In fact, we found a double transition for the total magnetization $m_T$. This behavior is due to the anti-ferromagnetic behavior introduced by negative values of the coupling exchange interactions ($J_{\sigma s} = J_{\sigma q} = -0.1$).

In Fig. 3d, we present the susceptibilities as a function of the crystal field in the absence of the external magnetic field (H = 0) and fixed values of the coupling exchange interactions: $J_{\sigma\sigma} = J_{ss} = J_{qq} = 1$ and $J_{\sigma s} = J_{\sigma q} = -0.1$. Indeed, only one peak is found for the susceptibilities $\chi_q$ and $\chi_T$. This peak corresponds exactly to the critical temperature of the system.

In order to show the effect of the behavior of the magnetizations as a function of the temperature, we plot in Figs. 4a and 4b the thermal magnetizations and susceptibilities profiles, respectively. This figure is plotted for the external magnetic field (H = 0), fixed values of the coupling exchange interactions $J_{\sigma\sigma} = J_{ss} = J_{qq} = 1$, $J_{\sigma s} = J_{\sigma q} = -0.1$ and D = $-1$. Indeed, we found three compensation temperature points for T < 2.5. While, the all magnetizations stabilize for T > 3, *see* Fig. 4a. In Fig. 4b, we provide the corresponding susceptibilities as a function of the temperature for (H = 0), fixed values of the coupling exchange interactions $J_{\sigma\sigma} = J_{ss} = J_{qq} = 1$, $J_{\sigma s} = J_{\sigma q} = -0.1$ and the crystal field value D = $-1$. All these susceptibilities show a peak, not only for the critical temperature Tc, but also at the temperature value corresponding to the compensation value, see Fig. 4b.

### 4- Conclusion:

In this work, we have studied the magnetic properties and phase diagrams of a Lieb nano-structure based on magnetic particles with «core-shell» consisting of ferromagnetic spins. We have examined the effect of the coupling exchange interactions and the crystal field. We also envisaged the stable configurations in different planes and found a perfect symmetry regarding the external magnetic field H . Some new phases are arising with large regions in the absence of the external magnetic field (H = 0). We applied the Monte Carlo simulations to study the behavior of this Lieb nano-structure for non null temperature values. On the other hand, we presented and discussed the hysteresis loops. We have also analyzed the effect of different physical parameters on the behavior of the partial and total magnetizations, susceptibilities, the compensation and critical temperatures. For specific values of the physical parameters, three compensation points are found for this Lieb nano-structure system.

**Figures captions:**

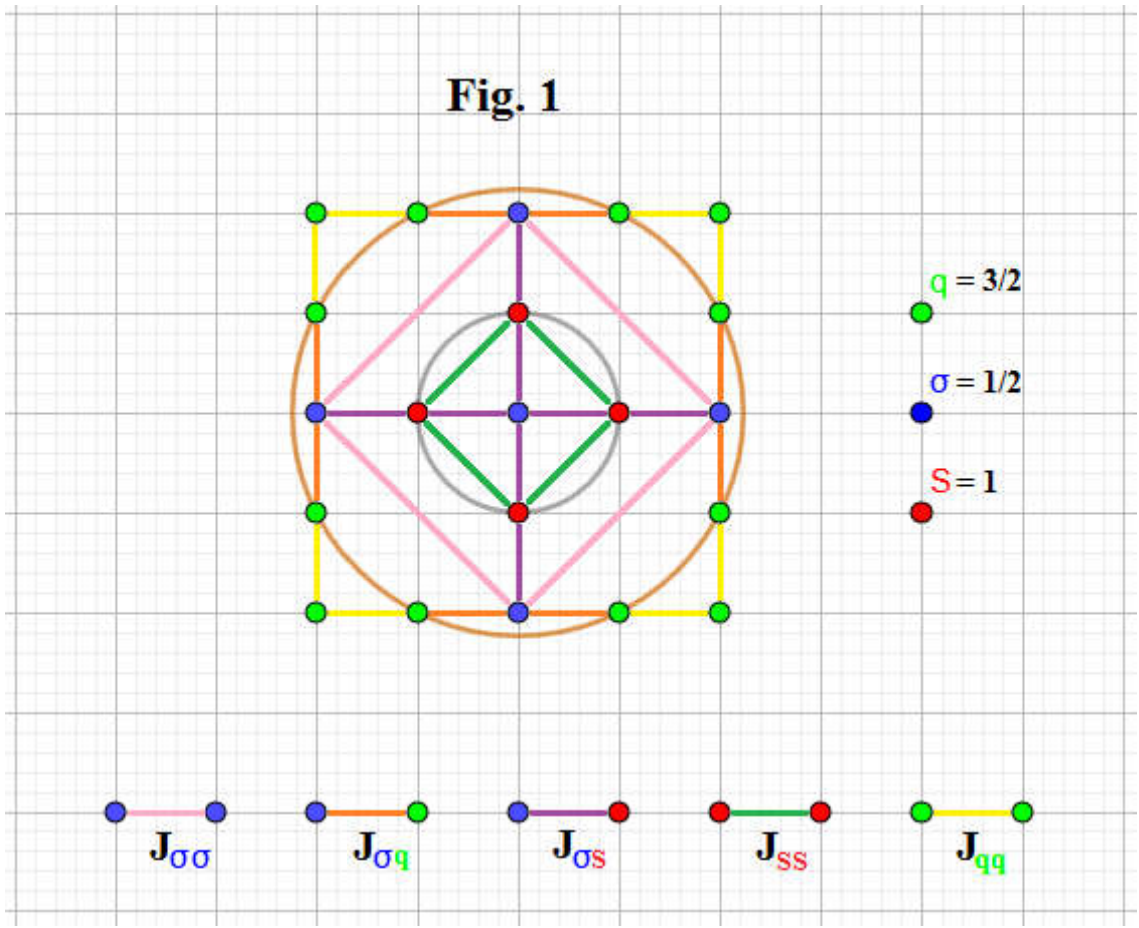

**Fig. 1:** Geometry of the studied system consisting of a core/shell with multiple spins $\sigma = 1/2$, $S = 1$ and $q = 3/2$, with: $N_\sigma = 5$, $N_S = 4$ and $N_q = 12$.

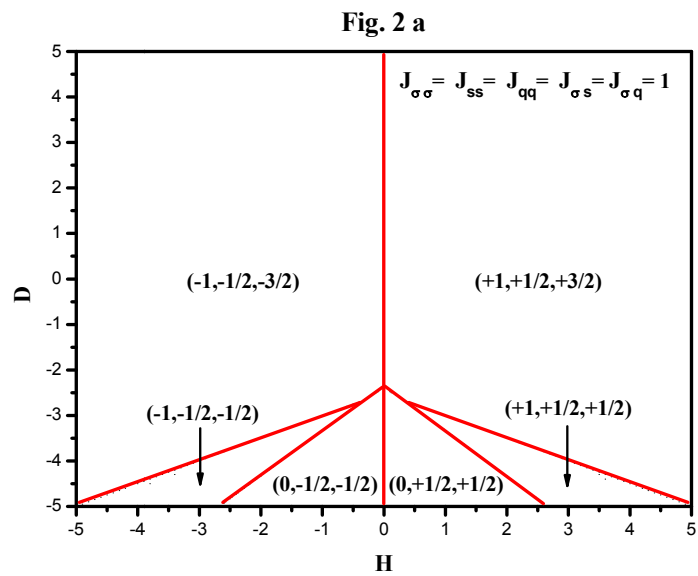

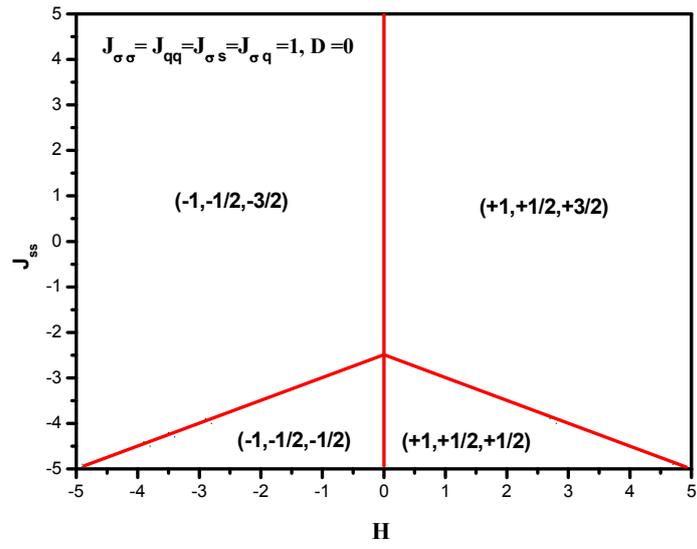

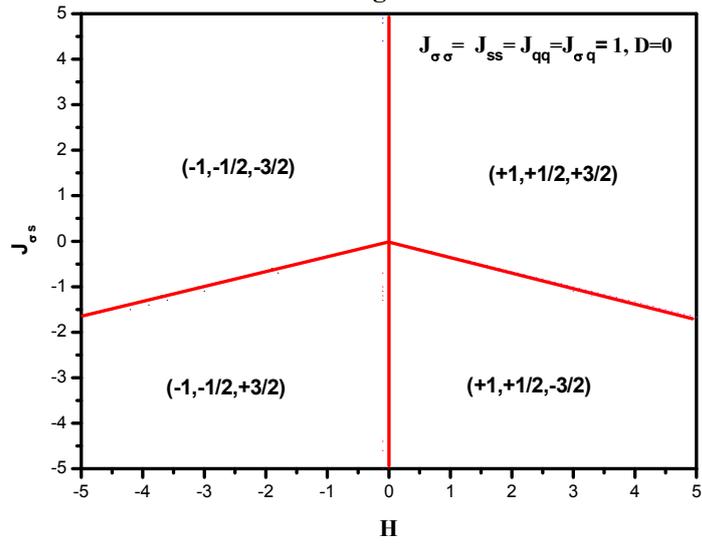

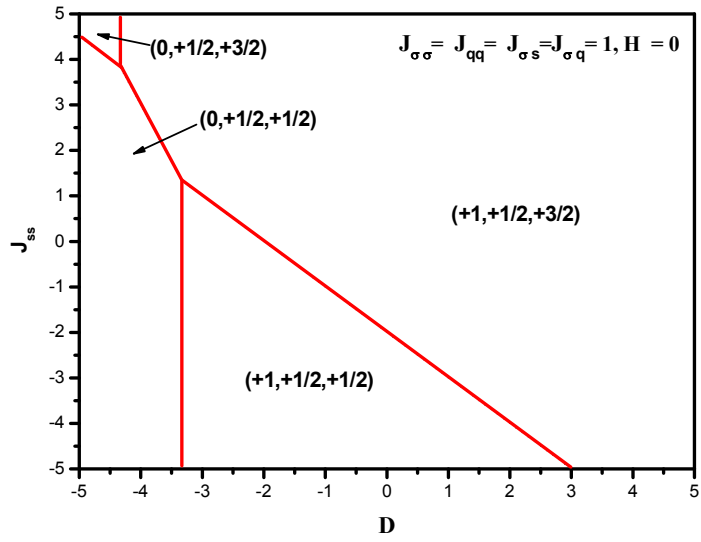

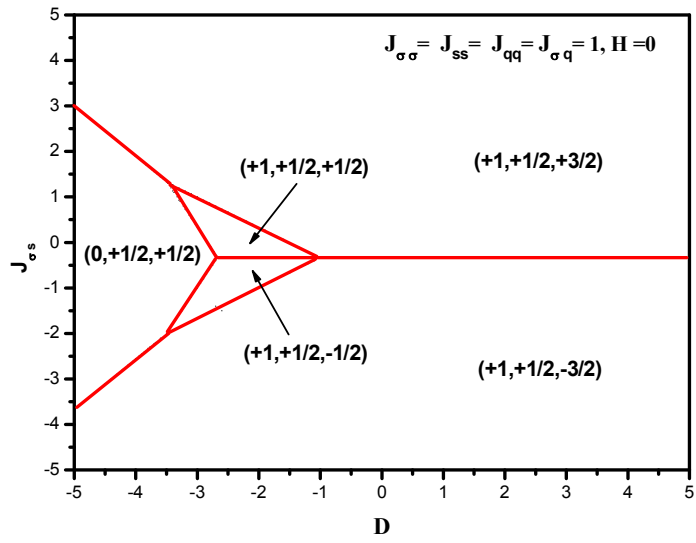

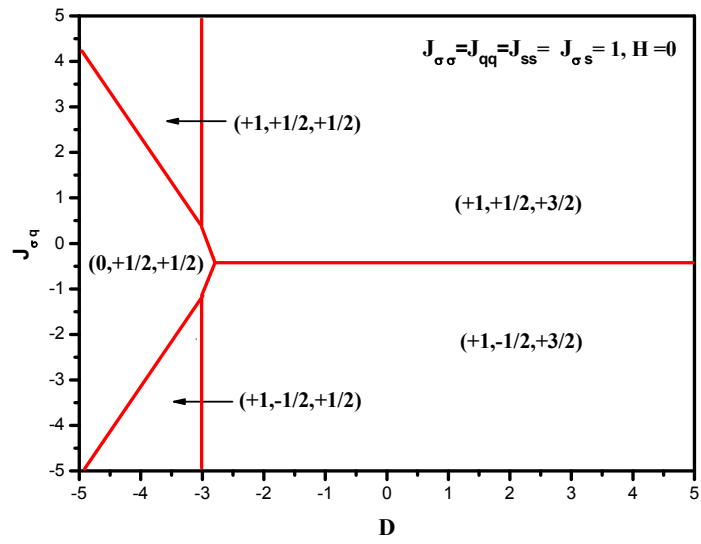

**Fig. 2 f**

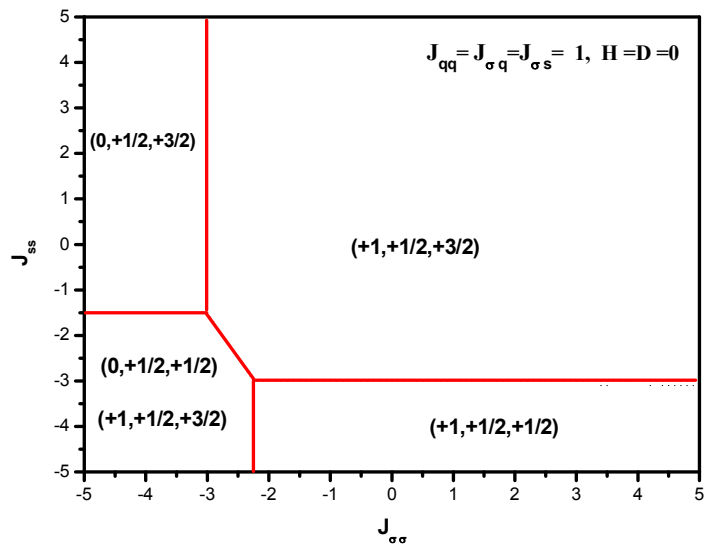

**Fig. 2 g**

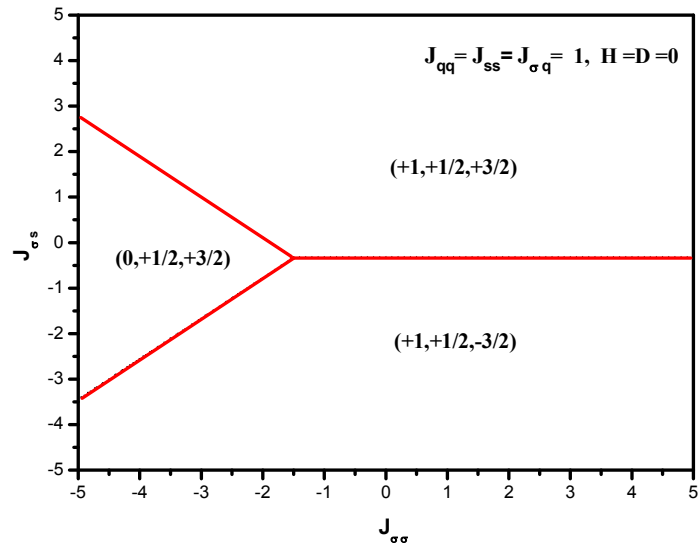

Fig. 2 h

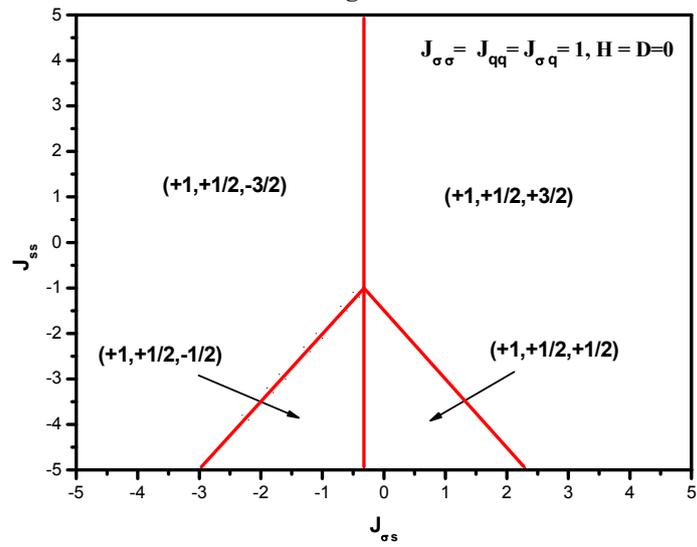

Fig. 2 i

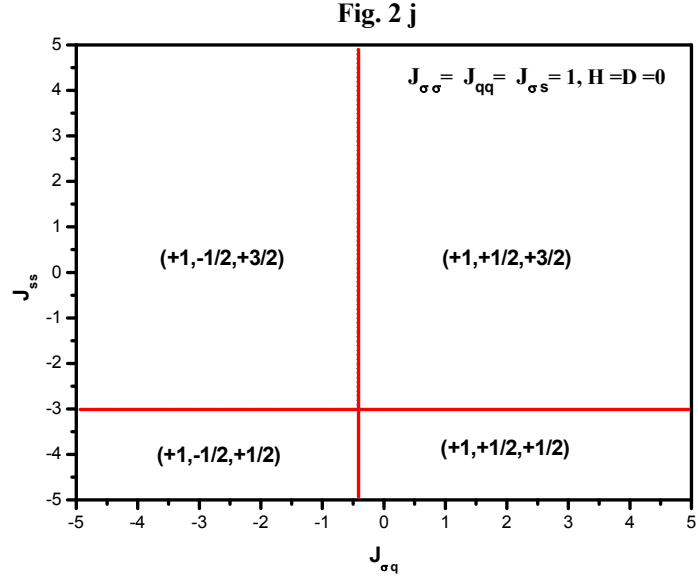

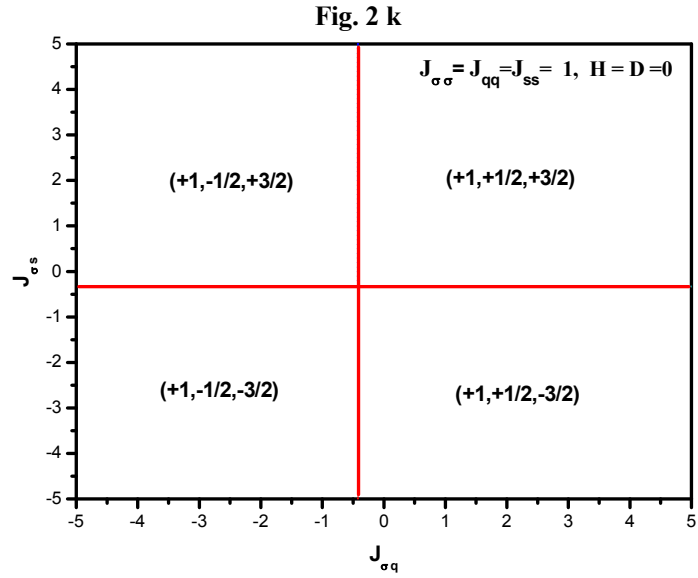

**Fig. 2:** Ground state phase diagrams of the Lieb magnetic structure showing the stable configurations:

(a) In the plane (H, D) for : $J_{\sigma\sigma} = J_{ss} = J_{qq} = J_{\sigma s} = J_{\sigma q} = 1$.

(b) In the plane (H, $J_{ss}$) for : $J_{\sigma\sigma} = J_{qq} = J_{\sigma s} = J_{\sigma q} = 1$ and $D = 0$.

(c) In the plane (H, $J_{\sigma s}$) for : $J_{\sigma\sigma} = J_{ss} = J_{qq} = J_{\sigma q} = 1$ and $D = 0$.

(d) In the plane (D, $J_{ss}$) for : $J_{\sigma\sigma} = J_{qq} = J_{\sigma s} = J_{\sigma q} = 1$ and $H = 0$.

(e) In the plane (D, $J_{\sigma s}$) for : $J_{\sigma\sigma} = J_{ss} = J_{qq} = J_{\sigma q} = 1$ and $H = 0$.

(f) In the plane (D, $J_{\sigma q}$) for : $J_{\sigma\sigma} = J_{ss} = J_{qq} = J_{\sigma s} = 1$ and $H = 0$.

(g) In the plane ($J_{\sigma\sigma}$, $J_{ss}$) for : $J_{qq} = J_{\sigma s} = J_{\sigma q} = 1$ and $H = D = 0$.

(h) In the plane ($J_{\sigma\sigma}$, $J_{\sigma s}$) for : $J_{ss} = J_{qq} = J_{\sigma q} = 1$ and $H = D = 0$.

(i) In the plane ($J_{\sigma s}$, $J_{ss}$) for : $J_{\sigma\sigma} = J_{qq} = J_{\sigma q} = 1$ and $H = D = 0$.

(j) In the plane ($J_{\sigma q}$, $J_{ss}$) for : $J_{\sigma\sigma} = J_{qq} = J_{\sigma s} = 1$ and $H = D = 0$.

(k) In the plane ($J_{\sigma q}$, $J_{\sigma s}$) for : $J_{\sigma\sigma} = J_{ss} = J_{qq} = 1$ and $H = D = 0$.

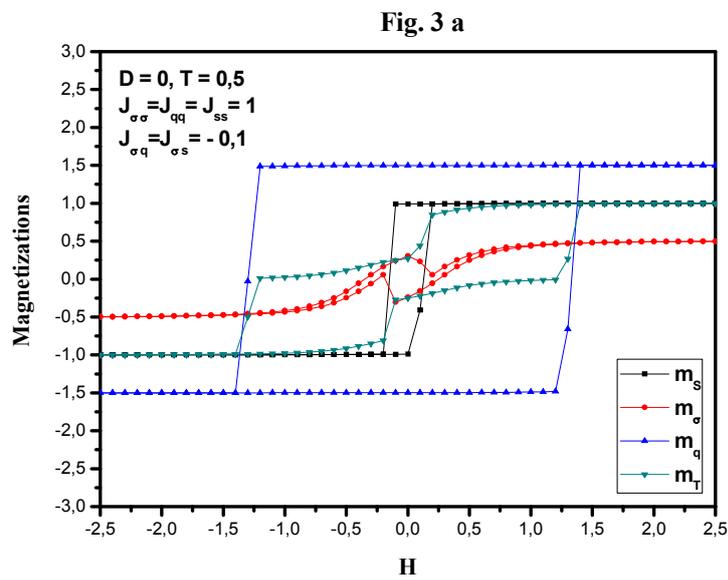

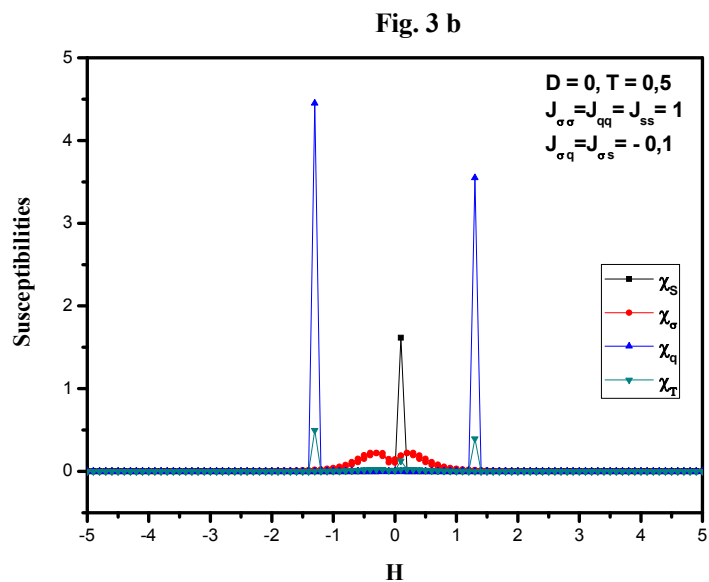

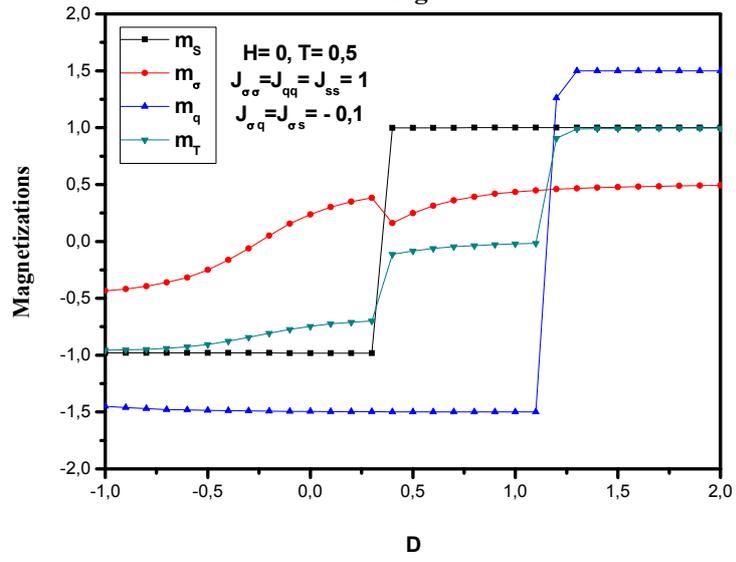

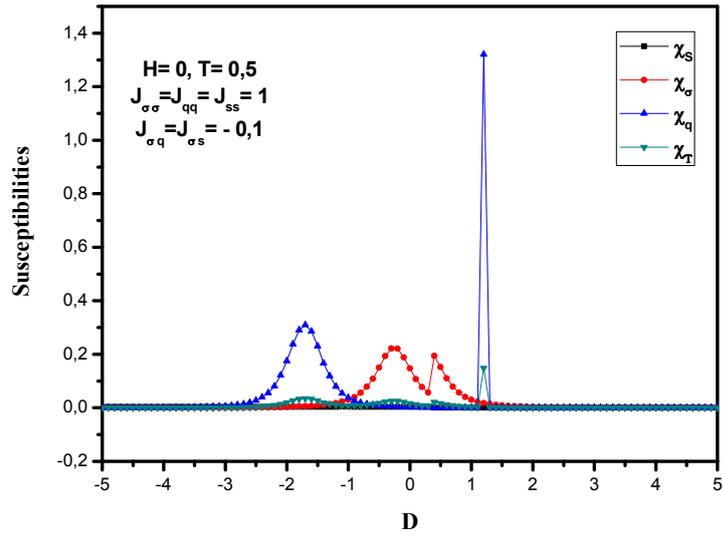

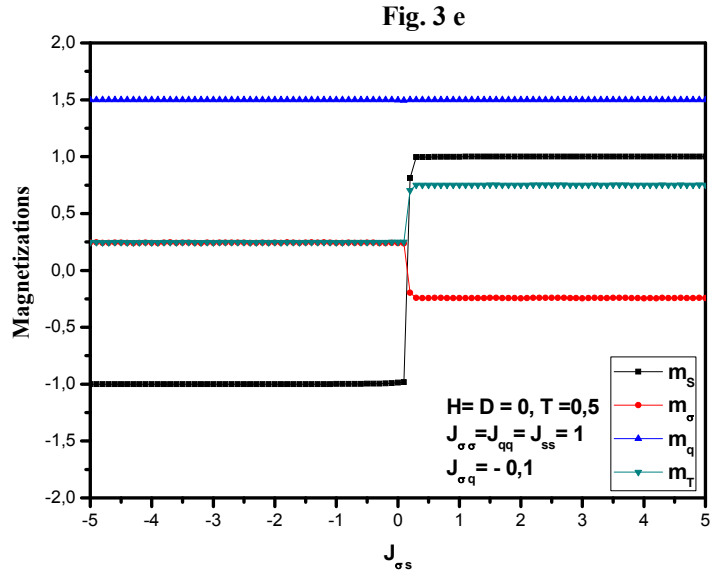

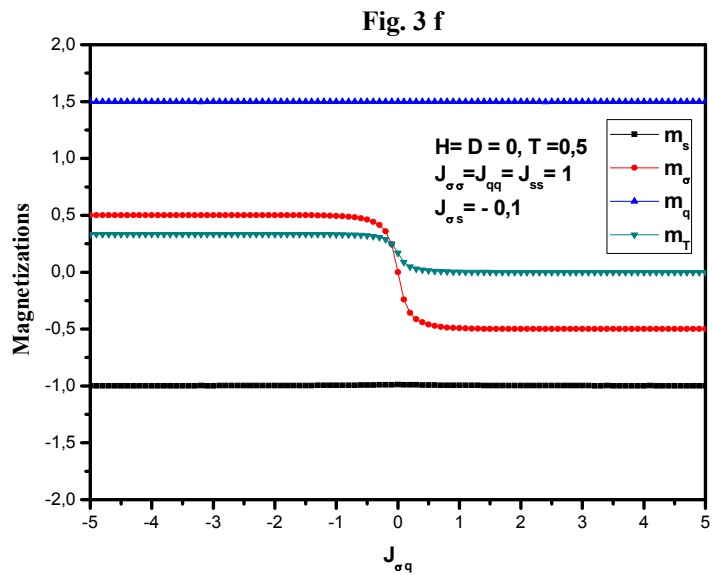

**Fig. 3:** The magnetizations and the susceptibilities profiles for: $J_{\sigma s} = -0.1$, $T = 0.5$, $H = D = 0$ and $J_{\sigma\sigma} = J_{ss} = J_{qq} = 1$. In (a) magnetizations and in (b) the susceptibilities profiles as a function of external magnetic field H, for $J_{\sigma\sigma} = J_{ss} = J_{qq} = 1$. In (c) the magnetizations and (d) the susceptibilities profiles as a function of crystal field D. The magnetization profiles as a function, in (e) of $J_{\sigma s}$ and in (f) $J_{\sigma q}$.

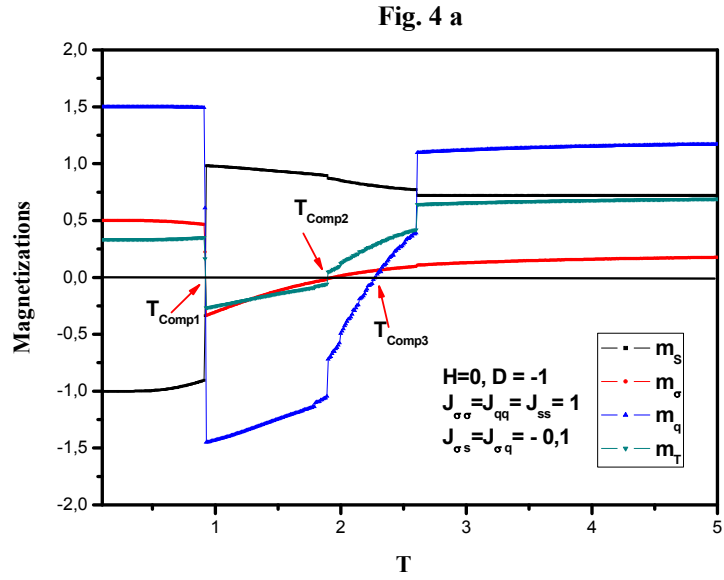

**Fig. 4:** The thermal magnetizations and susceptibilities profiles for: $J_{\sigma\sigma} = J_{ss} = J_{qq} = 1$ and $J_{\sigma s} = J_{\sigma q} = -0.1$, with: $H = 0$, $D = -1$. In (a) the magnetizations. In (b) the susceptibilities.